\documentclass[12pt]{article}
\usepackage{graphicx}
\begin{document}

\bigskip
\centerline{Bit-strings and other modifications of Viviane model for
language competition}

\bigskip

\noindent
P.M.C. de Oliveira$^{1,2}$, D. Stauffer$^{1,3}$, F.W.S. Lima$^4$, A.O. Sousa$^{1,5}$, 
C. Schulze$^3$ and S. Moss de Oliveira $^{1,2}$

\bigskip
\noindent
$^1$ Laboratoire PMMH, \'Ecole Sup\'erieure de Physique et de Chimie
Industrielles, 10 rue Vauquelin, F-75231 Paris, France

\medskip
\noindent
$^2$ Visiting from Instituto de F\'{\i}sica, Universidade
Federal Fluminense; Av. Litor\^{a}nea s/n, Boa Viagem,
Niter\'{o}i 24210-340, RJ, Brazil

\medskip
\noindent
$^3$ Inst. for Theoretical Physics, Cologne University, D-50923 K\"oln, Euroland

\medskip
\noindent
$^4$ Departamento de F\'{\i}sica,
Universidade Federal do Piau\'{\i}, 57072-970 Teresina - PI, Brazil

\medskip
\noindent
$^5$ Visiting from SUPRATECS, University of Li\`ege, B5, Sart Tilman, 
B-4000 Li\`ege, Euroland

\bigskip
Keywords: linguistics, Monte Carlo simulation, language size distribution

\bigskip
{\bf Abstract}

{\small The language competition model of Viviane de Oliveira et al is modified
by associating with each language a string of 32 bits. Whenever a language
changes in this Viviane model, also one randomly selected bit is flipped. 
If then only languages with different bit-strings are counted as different,
the resulting size distribution of languages agrees with the empirically 
observed slightly asymmetric log-normal distribution. Several other 
modifications were also tried but either had more free parameters or agreed
less well with reality.} 

\section{Introduction}

\begin{figure}[hbt]
\begin{center}
\includegraphics[angle=-90,scale=0.5]{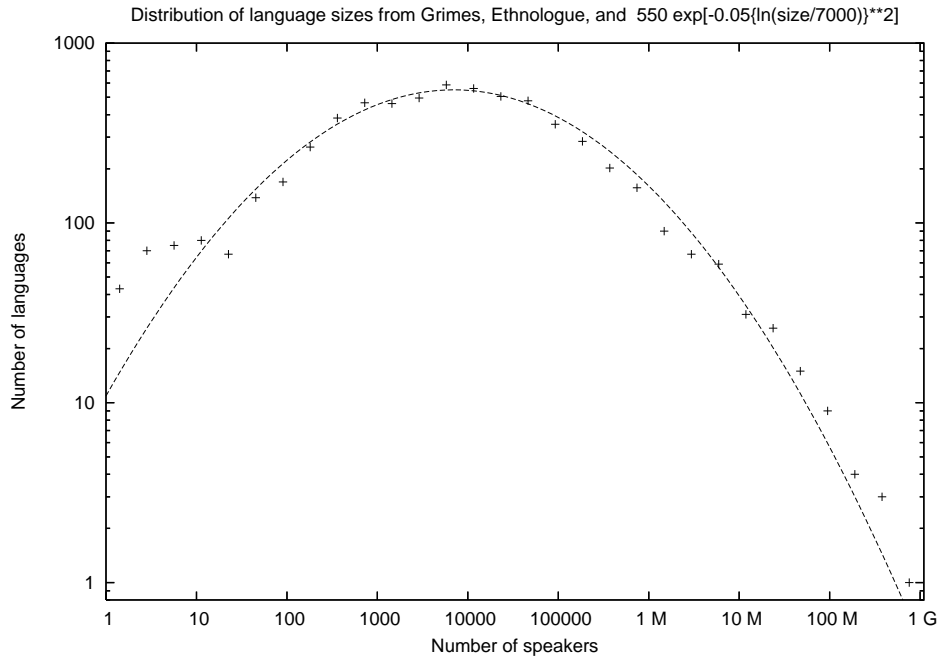}
\end{center}
\caption{Empirical size distribution of the $\sim 10^4$ present human languages,
binned in powers of two. The curve shows a fitted parabola, corresponding to 
a log-normal distribution. Real numbers of languages are for small languages
higher than this parabolic fit. From \cite{langsol}.
}
\end{figure}

The competition between languages of adult humans, leading to the extinction of 
some, the emergence of new and the modification of existing languages, has 
been simulated recently by many physicists [1-11]
and others [12-14], see also \cite{Nowak} for the learning
of languages by children. The web site http://www.isrl.uiuc.edu/amag/langev/ 
lists $10^3$ linguistic computer simulations, and recent reviews of language 
competition simulations were given in [16-18]. Perhaps the empirically
best-known aspect of language competition is the present distribution $n_s$ of 
language sizes $s$, where the size $s$ of the language is defined as the number
of people speaking mainly this language, and $n_s$ is the number of different 
languages spoken by $s$ people. We leave it to linguists and politicians to 
distinguish
languages from dialects and rely on the widely used ``Ethnologue'' statistics
 [19-22] repeated in Fig.1. This log-log 
plot shows a slightly asymmetric parabola, corresponding to a log-normal
distribution with enhancement for small sizes $s \sim 10$. Our aim is to 
reproduce this empirically observed distribution in an equilibrium simulation;
previously it was achieved only for non-equilibrium \cite{langsol}.

Of the many models cited above only the ``Schulze'' model \cite{Schulze}
and the ``Viviane'' model \cite{Oliveira} gave thousands of languages as
in reality. The Schulze model gave a reasonable $n_s$ distribution in 
non-equilibrium \cite{langsol}, when observed during its phase transition
between the dominance of one language spoken by most people and the 
fragmentation into numerous small languages. The Viviane model does not have 
such a phase transition \cite{cise}, and we now attempt to get from it a 
realistic $n_s$ in equilibrium.

The next section defines the standard Viviane model \cite{Oliveira} for the
reader's convenience. Section 3 gives our bit-string modification and the
improved resulting $n_s$, while Section 4 lists other attempts to get a good
size distribution. The concluding section 5 compares our various attempts.

\section{Viviane Model}

The original Viviane model \cite{Oliveira} simulates the spread of humans
over a previously uninhabited continent. Each site $j$ of an $L \times L$
square lattice can later be populated by $c_j$ people, where $c_j$ is initially
fixed randomly between 1 and $m \sim 10^2$. On a populated site only one 
language is spoken. Initially only one single site $i$ is occupied by $c_i$ 
people. 

Then as in Eden cluster growth or Leath percolation algorithm, at each time step
one surface site (= empty neighbour $j$ of the set of all occupied 
sites) is selected randomly, and then occupied with probability $c_j/m$ by
$c_j$ people. These settlers first select as language that of one of their 
occupied neighbour sites, with a probability proportional to the fitness of 
that language. This fitness $F_k$ is the total number of people speaking the 
language $k$ of that site, summed over all lattice sites occupied at that time.
(In \cite{Oliveira2}, this fitness was bounded from 
above by a maximum $M_k$ selected randomly between 1 and $M_{\max} \sim 20 m$.)
After a language is selected, it is mutated into a new language with probability
$\alpha/F_k$ with a mutation factor $\alpha$ typically between $10^{-3}$ and 
1. From then on the population and language of the just occupied lattice 
site remain constant. Equilibrium is reached when all lattice sites have 
become occupied and the simulation stops. As a result of this algorithm, the 
various languages are numbered 1, 2, 3, ... without any internal structure of 
the languages.

\begin{figure}[hbt]
\begin{center}
\includegraphics[angle=-90,scale=0.5]{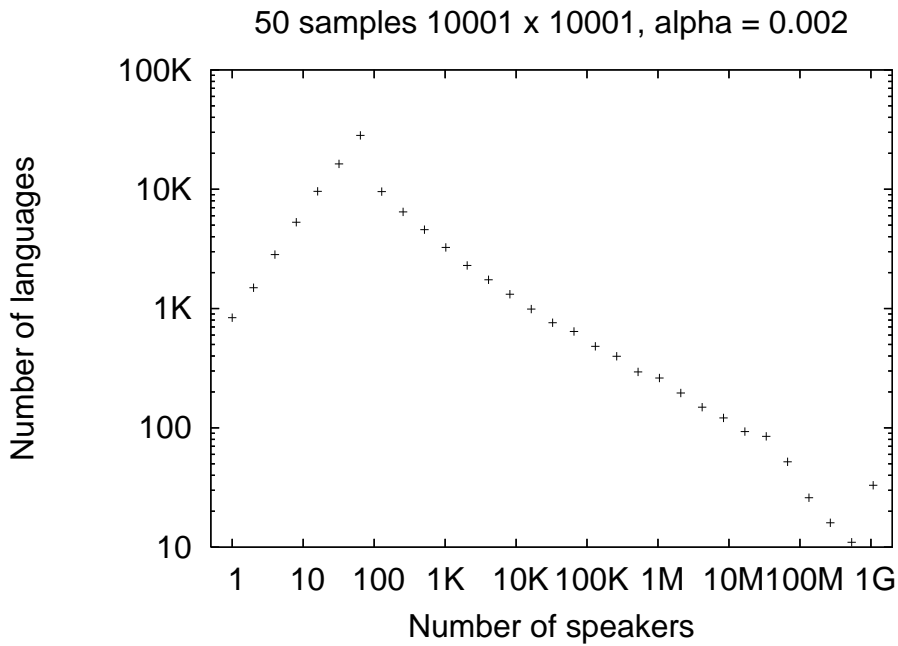}
\end{center}
\caption{Language size distribution $n_s$ for the standard Viviane model, with 
$s$ varying from 1 to $10^9$. The absolute value of the slope to the right is 
smaller than one on the left, in contrast to reality, Fig.1. 
$m = 127, \; M_{\max} = 16 m$,  also in Figs.3,5,6,8.
}
\end{figure}
 
\begin{figure}[hbt]
\begin{center}
\includegraphics[angle=-90,scale=0.5]{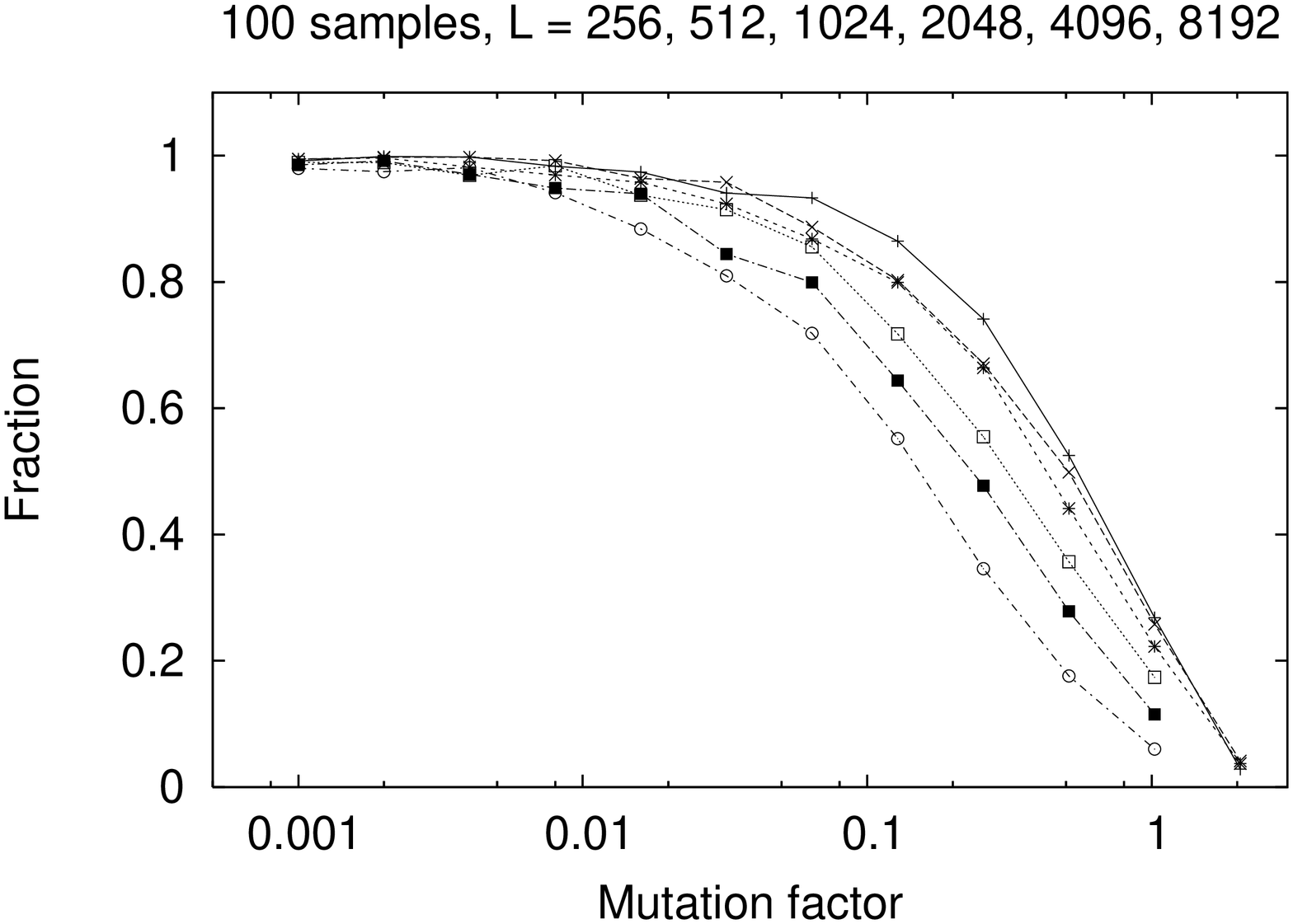}
\end{center}
\caption{Variation of the fraction of people speaking the largest language.
The linear lattice size $L$ increases from right to left.
For mutation factor $\alpha = 0$ by definition everybody speaks the language
of the initially occupied site.
}
\end{figure}
 
The resulting language size distribution $n_s$ in Fig.2 has a sharp maximum near
$s \sim m$, and follows one power law (exponent 1) to the left of the maximum
and another power law to its right. As in reality it extends from $s = 1$ to
$s = 10^9$ for the number $s$ of people speaking one language. But the 
sharp maximum is not seen in reality, Fig.1, and the simulated slope on the 
right of the maximum is weaker than the one at its left, while reality
shows the opposite asymmetry: Less slope on the left than on the right.

With increasing mutation factor $\alpha$, the fraction of people speaking the 
largest language decreases smoothly, Fig.3,  without showing a sharp phase 
transition (in contrast to the Schulze model). For increasing lattice size $L$ 
the curves shift slightly (logarithmically ?) to smaller $\alpha$ values.

(The program listed in \cite{cise} gave a limiting fitness $M_j$ to each site 
$j$, instead of an $M_k$ to each language. Thus before the mutations are 
simulated we need there the line {\tt f(lang(j))=min(limit(lang(j)), 
f(lang(j)) + c(j)*fac)}. This mistake barely affects the $n_s$, Fig.2, but after
correction the resulting size effect in our Fig.3 is weaker than in Fig.3 of
\cite{cise}.)

\section{Bit-string modification}

We now improve the Viviane model in three ways:

i)  We give the Viviane languages an internal structure by associating with each
language a string of, say, $\ell = 16$ bits, initially all set to zero. At each
mutation of the language at the newly occupied site, one randomly selected bit
is flipped, from 0 to 1 or from 1 to 0. We count languages as different only 
if they have different bit-strings. Otherwise the standard algorithm is 
unchanged. Thus our new bit-strings do not influence the dynamics of the 
population spread, only the counting of languages.

ii)
Thus far the populations $c_j$ per site $j$ were homogeneously distributed 
between 1 and $m$. In reality, there are more bad than good sites for human 
settlement. We approximate this effect by assuming that the values of $c$, 
to be scattered
between 1 and $m$, no longer are distributed with a constant probability but
with a probability proportional to $1/c$. 

iii) Instead of occupying one randomly selected surface site $i$ with 
probability proportional to $c_i$, we saved lots of computer time by 
selecting two such surface sites and occupying the one with the bigger $c$.
 
(As a minor improvement we counted a neighbour language only once if two or 
more neighbours of the just occupied site speak that language.)

\begin{figure}[hbt]
\begin{center}
\includegraphics[angle=-90,scale=0.5]{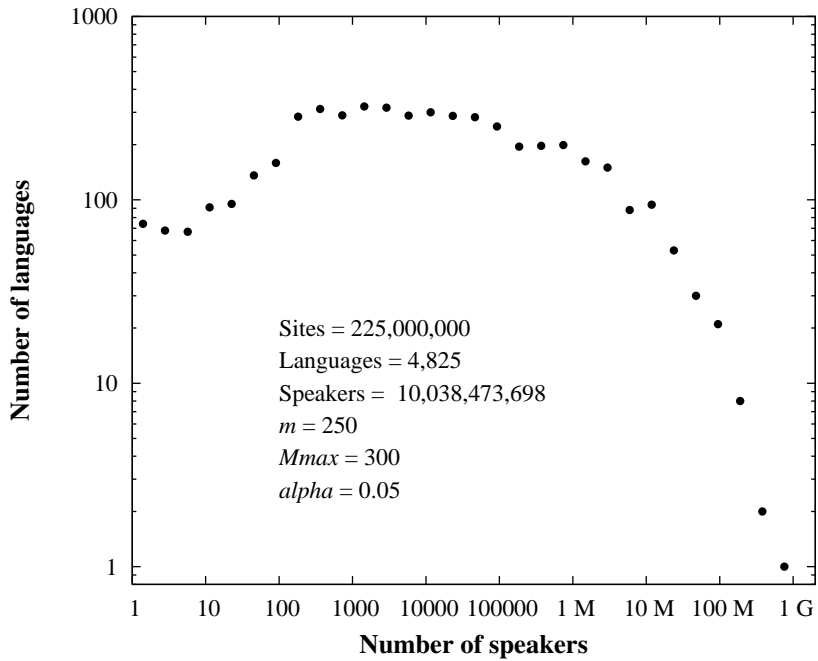}
\end{center}
\caption{Language size distribution for bit-string version, $L=15000, \; m=250, 
\; M_{\max}=300, \; \alpha = 0.05, \; \ell = 14$ bits.
}
\end{figure}
 
Fig.4 shows that these modifications are good enough to result in reasonable 
agreement with reality, Fig.1. The shape of the curve is robust against a wide 
variation of the parameters. We do not show plots for different $m$ since $1 
\le F_j \le m$ and for fixed $m/M_{\max}$ the simulations depend only on the 
ratio $\alpha/F_j$. The total number of languages is only $5 \times 10^3$,
less then the real \cite{Grimes} value $ 7 \times 10^3$ for which we would 
need bigger lattices than our computer memory can store.

As in \cite{Holman} for the Schulze model, the bit-strings allow a study
of spatial correlations: What is the Hamming distance for languages
separated by a distance $r$? The Hamming distance for two bit-strings, used 
already in \cite{Tesileanu,Holman} for the Schulze model, is the number of 
bits which differ from each other in a position-by-position comparison of the
two bit-strings. Thus initially we occupy the top line of the $L \times L$
lattice with $L$ different languages, all having bit-string zero, then
start the standard Viviane dynamics, and at the end we sum over all Hamming 
distances of all sites on lattice line $r$, compared with the corresponding
sites on the first lattice line. (By definition, this Hamming distance is zero
for $r = 1$.) Fig.5 shows our correlation functions, similar to reality
\cite{Holman,Goebl}; the higher the mutation factor $\alpha$, the higher
the Hamming distance. This simulation for Fig. 5 used only modification i) and 
involved no counting of languages.
 
\begin{figure}[hbt]
\begin{center}
\includegraphics[angle=-90,scale=0.30]{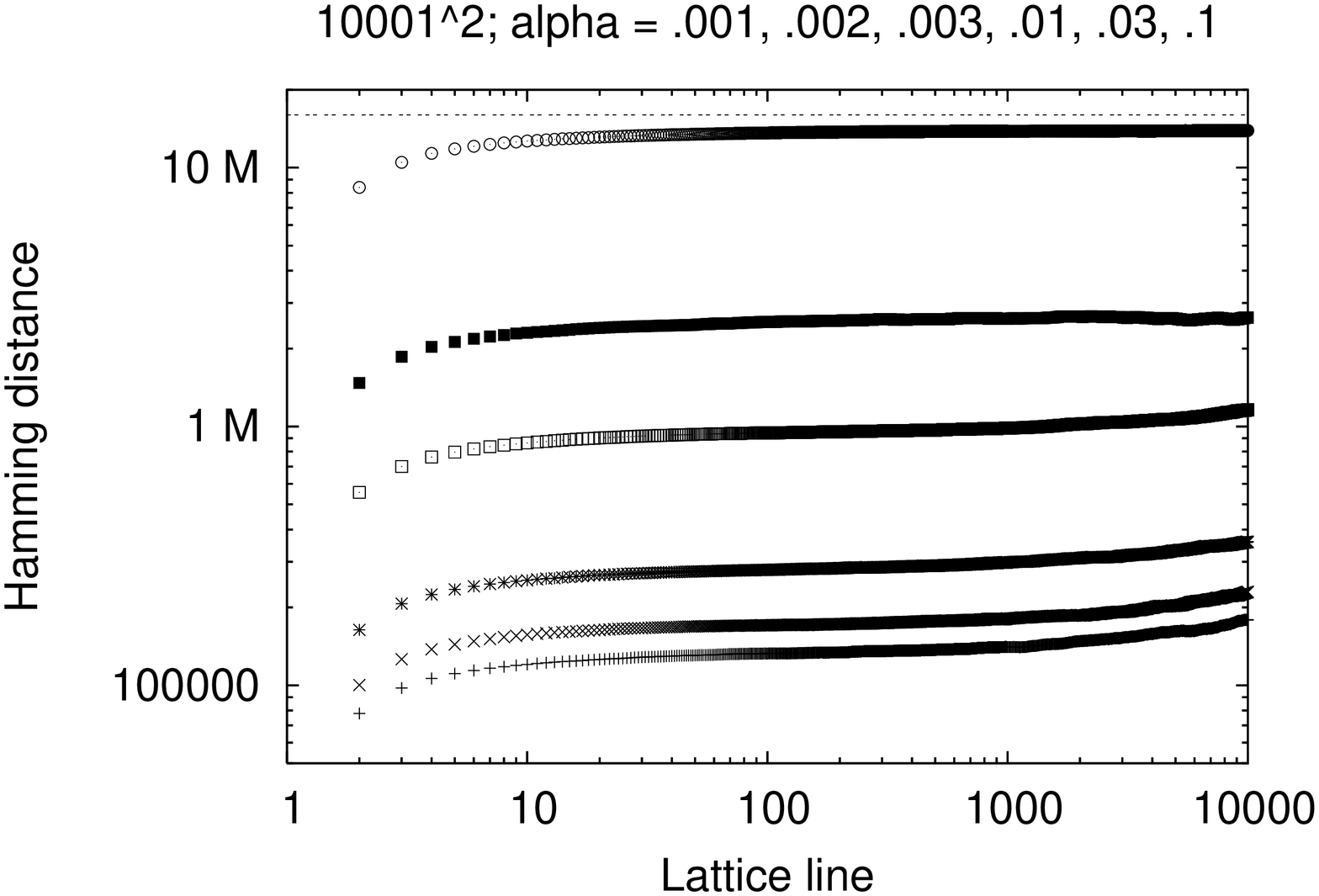}
\includegraphics[angle=-90,scale=0.30]{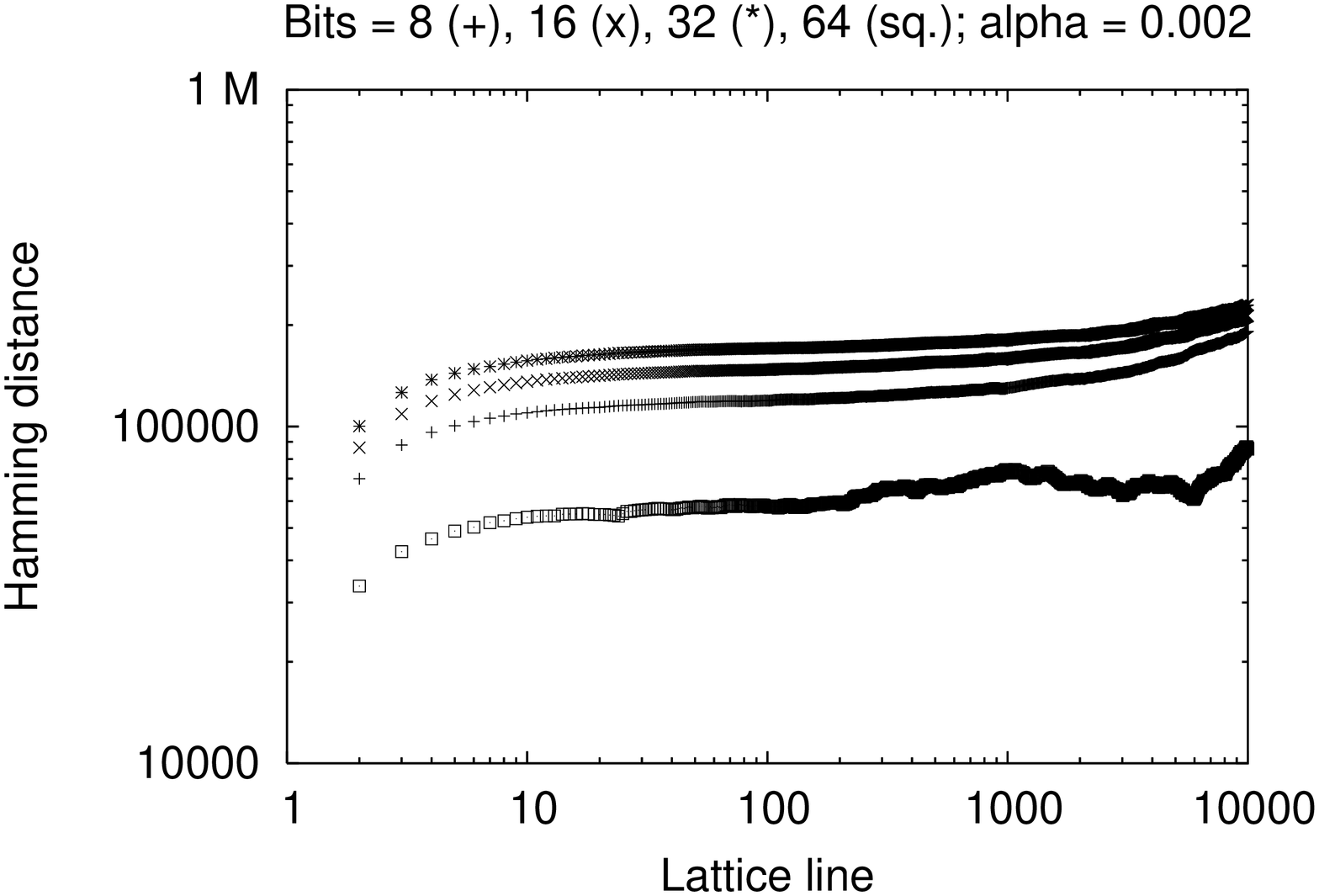}
\end{center}
\caption{Summed Hamming distance versus geometric distance. Upper part:
increase with increasing mutation factor, with the straight line on top 
giving the limit of uncorrelated bit-string. Lower part: variation with the 
length $\ell$ of our bit-string, taken as $\ell = 32$ in the upper part.
}
\end{figure}

\section{Other modifications}
\subsection{Noise}

Ref.\cite{langsol} improved the language size distribution of the Schulze model 
by applying random multiplicative noise, that means by multiplying at the
end of one simulation each $n_s$ repeatedly by a random number taken between
0.9 and 1.1. This modification approximates external influences from outside 
the basic model. Such noise is applied in Fig.6 to the standard Viviane 
model with the additional modification of correlations: each random
number is used twice, one after the other. Here we multiplied each $n_s$
thousand times  by a factor $(0.9+0.2z)^2$ at each iteration, and we summed 
over thousand samples. (Here $z$ is a random number homogeneously distributed 
between 0 and 1.) We start the simulations with a small mutation factor
$\alpha=0.001$ and for each iteration this grows linearly until it reaches 
a values of $\alpha=0.916$, for all lattices sizes used here: 
$L=257,513,1023,2047$ and 4095. Fig.6 shows a slightly asymmetric parabola, but
as in Fig.2 with the wrong asymmetry: Too slow decay on the right. 

\begin{figure}[hbt]
\begin{center}
\includegraphics[angle=-90,scale=0.5]{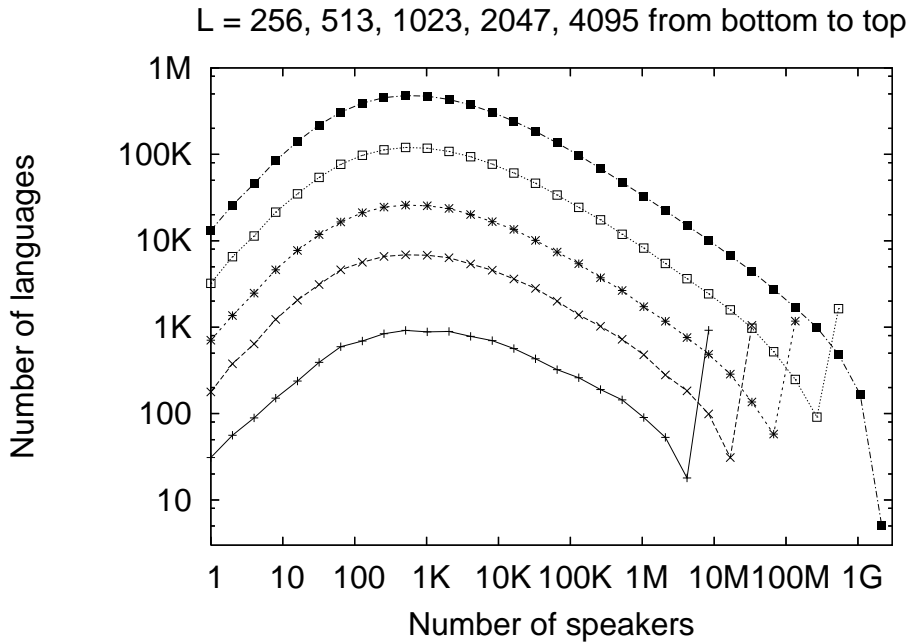}
\end{center}
\caption{Language size distribution from multiplicative noise and varying 
mutation factor (Viviane model without bit-strings).
}
\end{figure}

\subsection{Power law for populations per site}

Using only modification ii) of section 3, and
adding random multiplicative noise (100 multiplications with $0.9+0.2z$, without
correlations), Fig.7 now shows reasonable asymmetric parabolas for 
equilibrium, similar to \cite{langsol} for the non-equilibrium Schulze model.

\begin{figure}[hbt]
\begin{center}
\includegraphics[angle=-90,scale=0.5]{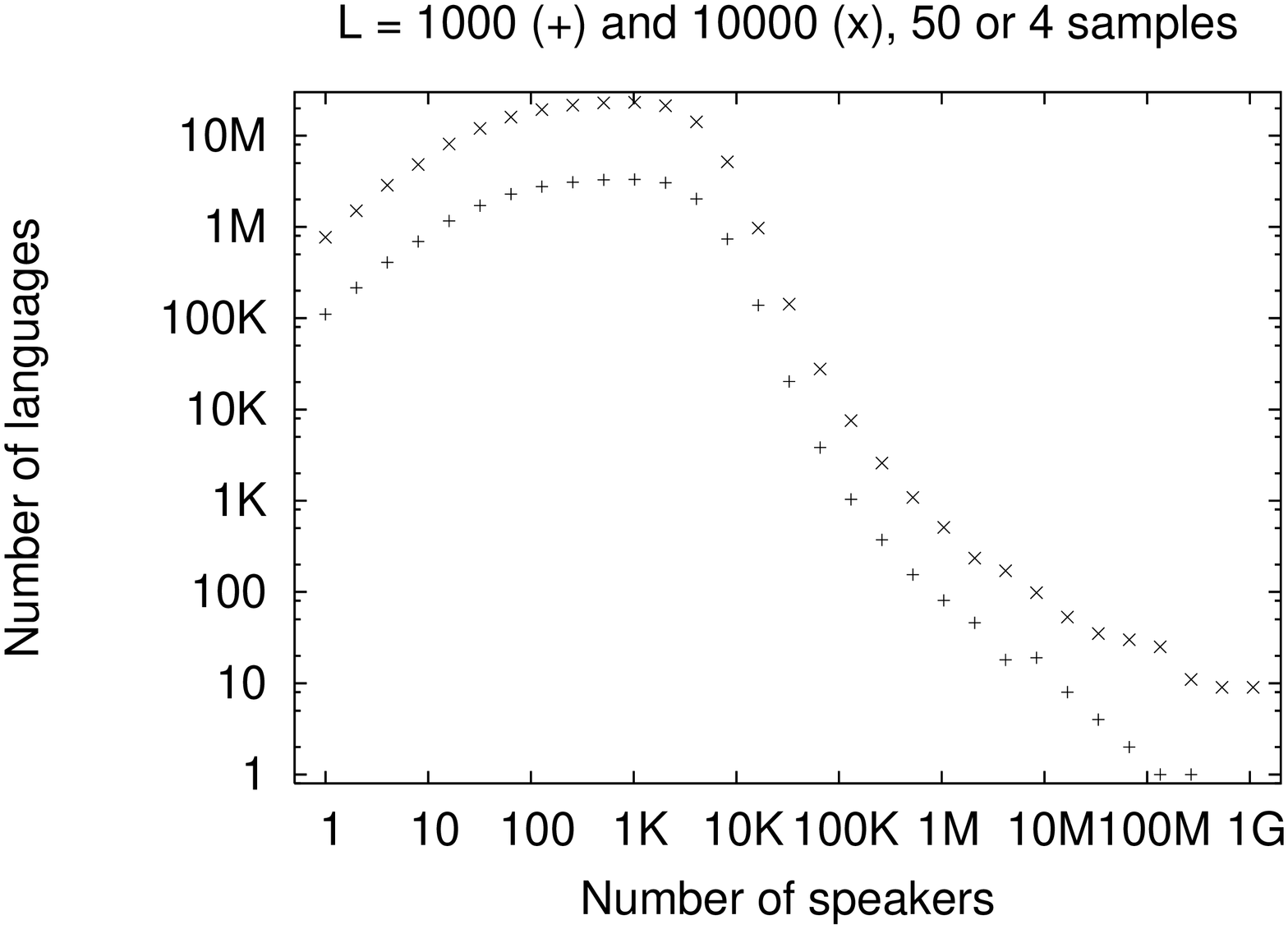}
\end{center}
\caption{Language size distribution with power law distribution for 
the $c_j$ and random multiplicative noise; $m = 8192, \; M_{max}=16m$ 
(Viviane model without bit-strings).
}
\end{figure}
 
\subsection{Indigenous population} 

We modified the standard Viviane model by assuming that initially
the lattice is not empty but is occupied by a native population
which in our simulation is then overrun by some foreign invaders.
Thus initially each lattice site gets a native fitness $1/z$ where $z$ is a 
random  number homogeneously distributed between zero and one. In the later
conquest by the foreign invaders, this site is conquered only if
the fitness of the invader is larger than the native fitness
(minus 10). It is possible that a few sites cannot be conquered,
since they are defended by Asterix, Obelix or other powerful natives.

We found that this modification barely changes the final
distribution of language sizes. For various mutation factors
$\alpha$, Fig. 8  shows that again we have two power
laws (straight lines in this log-log plot) for small and for large
language sizes. The time after which the ``conquistadores'' finish
their conquest varies very little from sample to sample (not shown). 
Adding as before random multiplicative noise by 100 multiplications by 
$0.9+0.2z$ makes the maximum more smooth (not shown), but still with the wrong
asymmetry.

\begin{figure}[hbt]
\begin{center}
\includegraphics[angle=-90,scale=0.5]{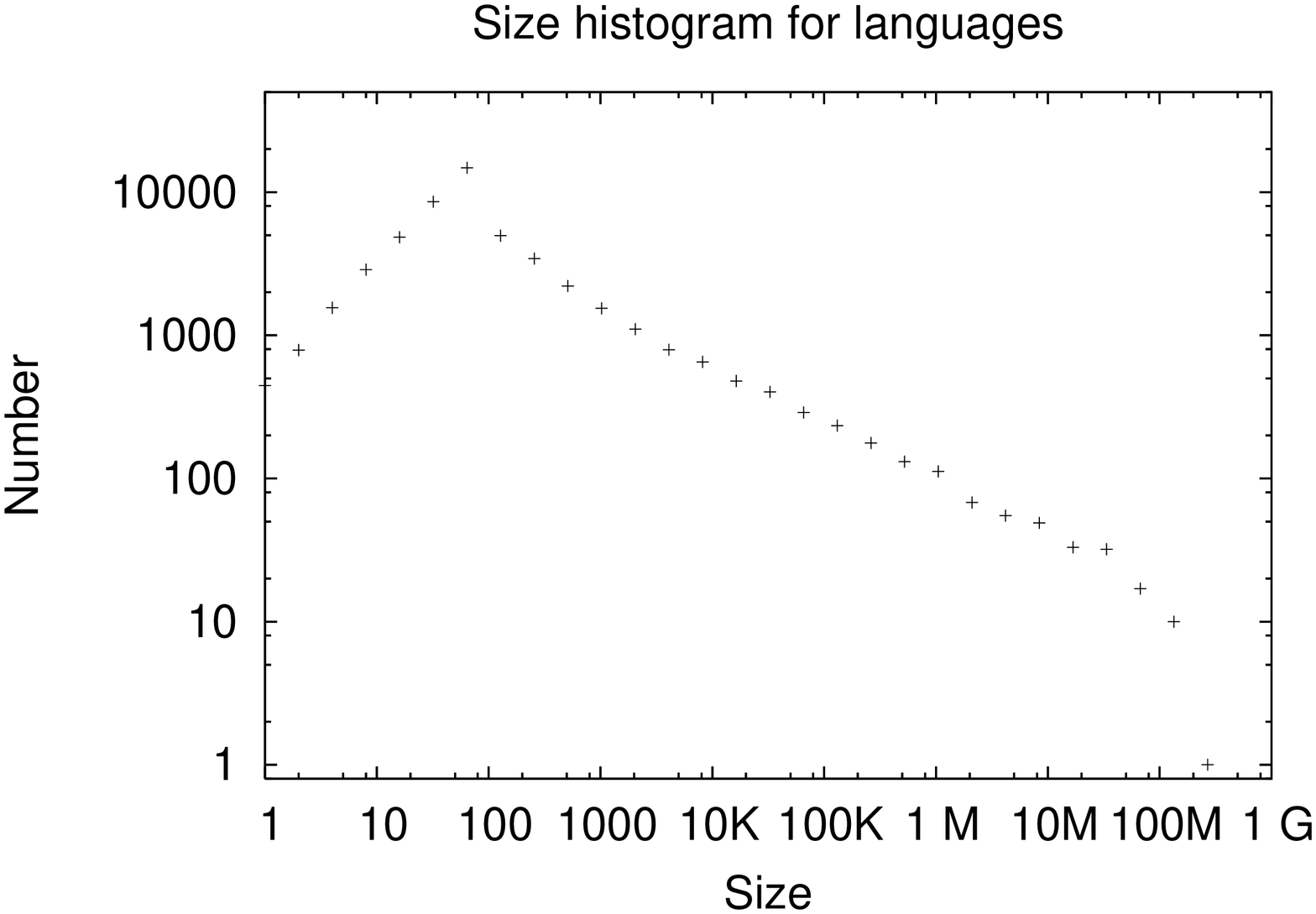}
\end{center}
\caption{Results similar to Fig.2 but with a native population at the beginning
of the conquest.
}
\end{figure}

\section{Conclusion}

While we have offered various modifications in order to improve the results 
from the standard Viviane model, we think the one of section 3 is the best since
it is simple and introduced no new free parameters except $\ell$. We have seen a
reasonable agreement with the slightly asymmetric log-normal distribution of 
language sizes. Future work could replace the bits by integer variables between 
1 and $Q$ as in some Schulze models \cite{cise}, or look at language families 
\cite{langwichmann}.

We thank the Brazilian grants PRONEX-CNPq-FAPERJ/171.168-2003 for financial 
support and S. Wichmann for many discussions.


\begin{thebibliography}{99}

\bibitem{Abrams}
D.M. Abrams and S.H. Strogatz, Nature 424 (2003) 900.

\bibitem{Patriarca}
M. Patriarca and T. Leppanen, Physica A 338 (2004) 296.

\bibitem{Mira}
J.Mira and A. Paredes, Europhys. Lett. 69 (2005) 1031.

\bibitem{Schulze}
C. Schulze and D. Stauffer, Int. J. Mod. Phys. C 16 (2005) 781 

\bibitem{Kosmidis}
K. Kosmidis, J.M. Halley and P. Argyrakis, Physica A 353 (2005) 595. 

\bibitem{Pinasco}
J.P. Pinasco and L. Romanelli, Physica A 361 (2006) 355; 

\bibitem{Schwammle}
V. Schw\"ammle, Int. J. Mod. Phys. C 16 (2005) 1519.

\bibitem{Oliveira}
V.M. de Oliveira, M.A.F. Gomes and I.R. Tsang, Physica A 361 (2006) 361

\bibitem{Oliveira2}
V.M. de Oliveira, P.R.A. Campos, M.A.F. Gomes and I.R. Tsang, Physica A 368 
(2006) 257.

\bibitem{Baronchelli}
A. Baronchelli, M. Felici, E. Caglioti, V. Loreto, L. Steels, 2006,
Sharp transition towards vocabularies in multi-agent systems, preprint.

\bibitem{mallorca} D. Stauffer, X. Costello, V.M. Egu\'{\i}luz and M. San 
Miguel, e-print physics/0603042 at www.arXiv.org for Physica A.

\bibitem{Nettle}
D. Nettle, Proc. Natl. Acad. Sci. 96 (1999) 3325.

\bibitem{Wang}
W.S.Y. Wang and J.W. Minett, Trans. Philological Soc. 103 (2005) 121.

\bibitem{Cangelosi}
A. Cangelosi and D. Parisi, eds., {\it Simulating the Evolution
of Language}, Springer, New York 2002.

\bibitem{Nowak}
M.A. Nowak, N.L. Komarova and P. Niyogi, Nature 417 (2002) 611.

\bibitem{book} D. Stauffer, S. Moss de Oliveira, P.M.C. de Oliveira, J.S. 
S\'a Martins, {\it Biology, Sociology, Geology by Computational Physicists},
Elsevier, Amsterdam 2006.

\bibitem{cise}
C. Schulze and D. Stauffer, Computing Sci. Engin. 8 (May/June 2006) 86

\bibitem{chachacha}
C. Schulze and D. Stauffer, page 307 in:  Econophysics \& Sociophysics: Trends 
\& Perspectives, eds: 
B.K.  Chakrabarti, A. Chakraborti and A. Chatterjee, Wiley-VCH, Weinheim 2006. 

\bibitem{Grimes} B.F. Grimes, 2000, {\it Ethnologue: languages of the world}  
(14th edn. 2000). Dallas, TX: Summer Institute of Linguistics; www.sil.org.

\bibitem{Sutherland} W.J. Sutherland, Nature 423 (2003) 276.

\bibitem{Gomes} M.A.F. Gomes, G. L. Vasconcelos, I. J. Tsang, and I. R. Tsang,
Physica A 271 (1999) 489.

\bibitem{Wichmann} S. Wichmann, J. Linguistics 41 (2005) 117.

\bibitem{langsol} D. Stauffer, C. Schulze, F.W.S. Lima, S. Wichmann and S.
Solomon, Physica A in press, physics/0601160 at arXiv.org.

\bibitem{Holman} 
E.W. Holman, C. Schulze, D. Stauffer \& S. Wichmann, physics/0607031 at arXiv.org.

\bibitem{Tesileanu}
T. Te\c sileanu and H. Meyer-Ortmanns, Int. J. Mod. Phys. C 17 (2006) 259.

\bibitem{Goebl} H. Goebl, H., Mitt. \"Osterr. Geogr. Ges. 146 (2004) 247.

\bibitem{langwichmann} S. Wichmann, D. Stauffer, F. W. S. Lima and C. Schulze. 
submitted to Transactions of the Philological Society, physics/0604146 at
arXiv.org.

\end{thebibliography}
\end{document}